\documentclass[a4paper,11pt,oneside]{article}
\usepackage{graphicx}				
\usepackage{listings}
\usepackage[colorlinks=true,linkcolor=green,anchorcolor=green,citecolor=green,filecolor=green,menucolor=green,urlcolor=blue,breaklinks=true,pdfhighlight=/P,pdfmenubar=true,pdftoolbar=true,pdfpagelabels=true,pdfstartpage=1,pdfstartview=FitV,pdftitle={Simulation of Pedestrians Crossing a Street},pdfsubject={Simulation of Pedestrians Crossing a Street},pdfauthor={Tobias Kretz},pdfcreator={Tobias Kretz},pdfproducer={Tobias Kretz},pdfkeywords={Pedestrian Traffic, Crowd Dynamics, Crossings, Simulation, VISSIM}]{hyperref}	
\usepackage[numbers,sort&compress]{natbib}	
\usepackage{hypernat}			
\usepackage[american]{babel}		
\begin{document}
%
\title{Simulation of Pedestrians Crossing a Street}
\author{Cornelia B{\"o}nisch and Tobias Kretz \\ \\
PTV Planung Transport Verkehr AG\\ 
Stumpfstra{\ss}e 1\\ 
D-76131 Karlsruhe\\ \\
\tt{\{Cornelia.Boenisch,Tobias.Kretz\}@ptv.de}}

\maketitle

\begin{abstract}
The simulation of vehicular traffic as well as pedestrian dynamics meanwhile both have a decades long history. The success of this conference series, PED and others show that the interest in these topics is still strongly increasing. This contribution deals with a combination of both systems: pedestrians crossing a street. In a VISSIM simulation for varying demand jam sizes of vehicles as well as pedestrians and the travel times of the pedestrians are measured and compared. The study is considered as a study of VISSIM's conflict area functionality as such, as there is no empirical data available to use for calibration issues. Above a vehicle demand threshold the results show a non-monotonic dependence of pedestrians' travel time on pedestrian demand.
\end{abstract}

\section{Introduction}
For vehicles and pedestrians alike the single mode systems have attracted first and much interest: highway traffic for vehicles \cite{Nagel1996,Chowdhury2000,Helbing2001b} and evacuations for pedestrians \cite{Schadschneider2009b,Schadschneider2009,Helbing2009}. These systems respectively situations are comparatively easy to handle in analytical or numerical terms and of special interest as they are most present in public awareness.

During the post-war re-building period of European cities the focus of city planning was strongly set to vehicular traffic. But recognizing economic, ecologic and aesthetic needs, the needs of pedestrians and cyclists were gaining ground. This shift in paradigms becomes most visible in the idea of and discussion on ``shared space'' and cities like Copenhagen that have explicitly put pedestrians into the top priority position.

In parallel to this development the rapid progress in computation hardware within just three decades made it possible to advance from simplified vehicular traffic simulations with limited extend to large scale combined simulations of vehicular traffic, cyclists, and pedestrians.

Recently VISSIM was the first professional tool to incorporate the simulation of vehicles and pedestrians as well as zones of interaction between these modes of traffic \cite{VISSIM2008}.

The amount of work done on ``interaction issues'' -- be it on signalized or non-signalized crossings -- is still marginal compared to the work done in the separate fields \cite{Tanner1951,Hunt1993,Vogts2001,Jiang2002,Lehnhoff2004,Helbing2005,Ishaque2007,Ishaque2008}.

\begin{figure}[htbp]
  \center
	\includegraphics[width=0.618\textwidth]{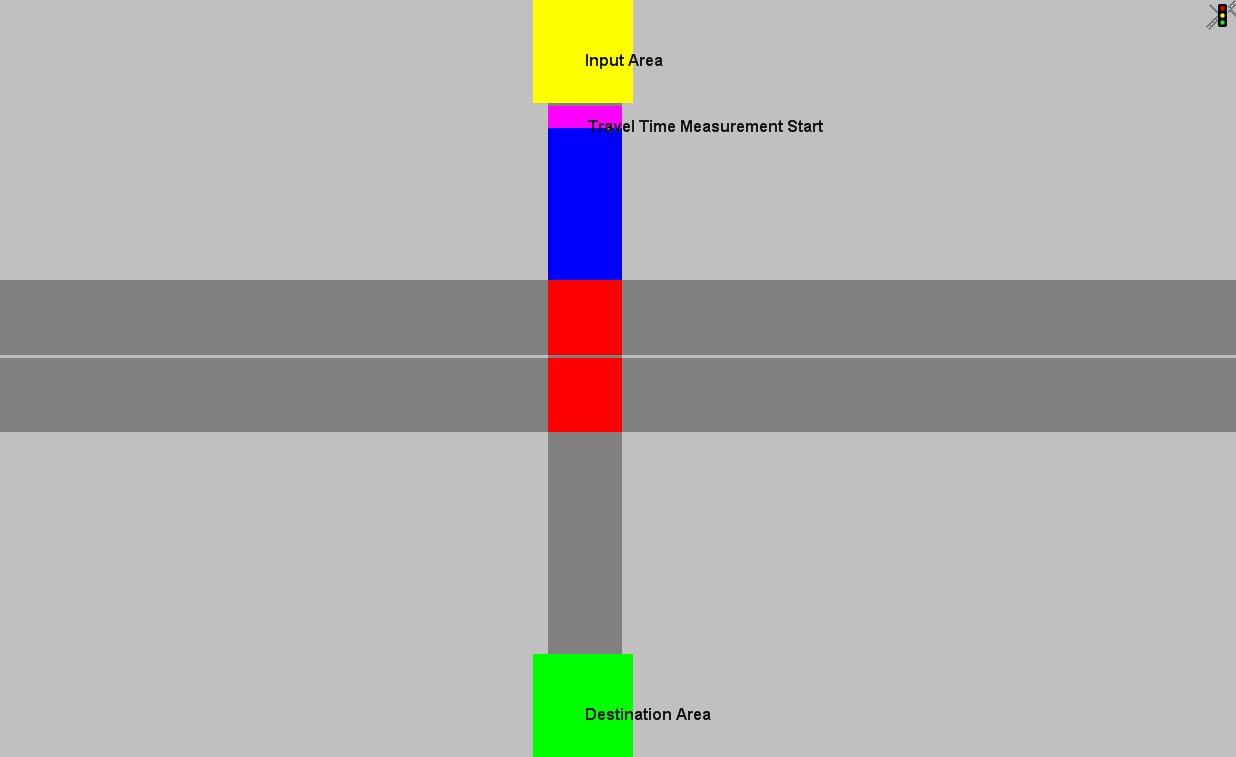}
	\caption{Pedestrians are inserted with a given average frequency to the simulation on the yellow area. Travel time measurement starts as soon as they are on the magenta area. They are counted as ``jam'' while they are either on the blue or the magenta area. The conflict areas are depicted red and each have an area of 3.5 x 3.5 sqm. Once a pedestrian has reached the green area, he's taken out of the simulation. The distance for the travel time measurement is 26 m. The vehicle lanes stretch 500 m to both sides. (They are not shown in their entirety here.)}
	\label{fig:geometry}
\end{figure}

In this work pedestrians crossing a street with a lane for each direction are simulated in VISSIM. The pedestrians only walk one-way, there is no counterflow. For the simulation of vehicles and pedestrians VISSIM's standard models are applied \cite{Wiedemann1974,Helbing1995,Helbing2000b,Johansson2007,VISSIM2008,Helbing2009}. One can expect that the underlying operational models only have a marginal influence on the results\footnote{It has to be considered in the construction of the conflict area that the acceleration in the Social Force Model is finite, if $\tau$ is larger than the simulation time step.}. In each simulation 10 hours were simulated. Demand was kept constant within a simulation and varied between simulations. For details of the geometry see figure \ref{fig:geometry}.

A note on demand and input areas: the maximum density on input areas at which still pedestrians are inserted to the simulation is 5 pedestrians per square meter. Even without any vehicles at all, this value is reached for input values of slightly less than 12,000 pedestrians per hour (in this case about 1,000 pedestrians in ten hours are skipped at input). Increasing the input further only has statistical effects on the simulation result.

\section{Conflict Areas}

The interaction zone is modeled as a ``conflict area'' similar to the way vehicle-vehicle conflicts were modeled earlier in VISSIM \cite{VISSIM2007}. A conflict area for vehicle-vehicle conflicts is an area, because vehicles have a width. In an abstract representation, it would be a conflict point, as two basically one-dimensional objects -- the links -- intersect. A conflict area for pedestrian-vehicle interaction is an area, because the pedestrians -- although compared to vehicles almost point-like -- can also have a transversal component in their movement compared to the main direction of motion, and the available passage width pedestrians have at crossings is usually a multiple of their body extension, which is different for vehicles, which on one lane have almost no transversal freedom.

Once a conflict area is defined, the priority is given either to vehicles or to pedestrians. This study deals exclusively with conflict areas with vehicle priority, i.e. a normal part of a street, no pedestrian crossing or even signalisation.

When approaching a conflict area, pedestrians calculate, if there is enough time to cross the street in time before the next vehicle arrives. But if the density of pedestrians is sufficiently high, pedestrians may be forced to slow down, or evade other pedestrians (move transversally and by this reduce the lateral speed). In this case, it may happen that they do not make it in time to the other side of the street. The information of pedestrians being on the conflict area is then given to the approaching vehicles, which in turn slow down, notwithstanding their right of way (an animation of this can be found online \cite{VISSIM2008}).

Pedestrians approaching from behind to the conflict area do not always base their decision to walk or to keep standing at the edge of the road on the vehicle's speed, but -- if there is at least one pedestrian on the conflict area -- they estimate, if they could overtake pedestrians on the conflict area with their desired speed. 

VISSIM's conflict areas have five parameters, whose values can have an effect in the situation discussed in this paper. They were not part of the investigation and thus set to equal values for all simulations. The visibility of both links was 100 m, the front and rear gap 0.5 s and the safety distance factor 1.5.

The usage of these parameters for conflicts between vehicles is fully described elsewhere \cite{VISSIM2008}, discussing it here, would exceed the size of this contribution. For the investigated situation where pedestrians are crossing a vehicle link the parameter for the rear gap is the most important one.
The rear gap for pedestrians has the same meaning like for vehicles, i.e. it is the time gap in seconds after the pedestrian left the conflict area and before the next vehicle enters.

\begin{figure}[htbp]
  \center
	\includegraphics[width=0.618\textwidth]{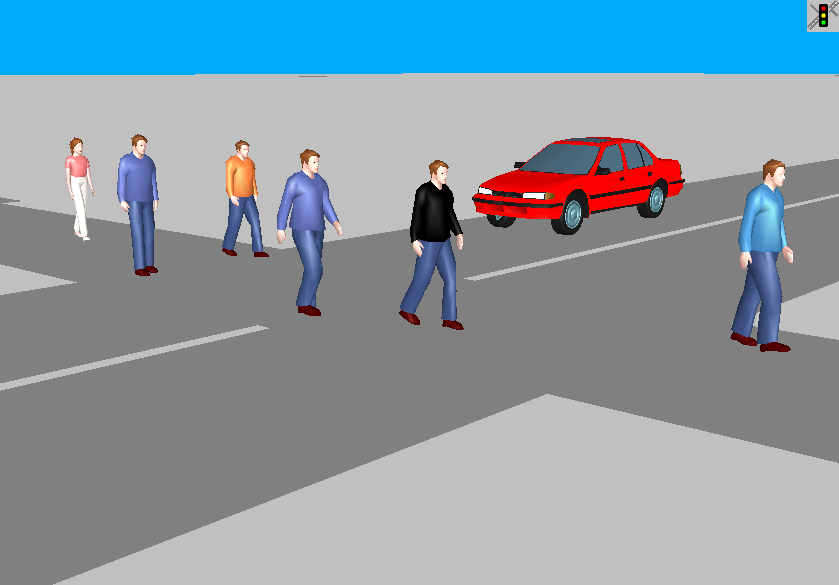}
	\caption{A screenshot from a 3d animation of the simulation (animation online \cite{VISSIM2008}).}
	\label{fig:3d}
\end{figure}

\section{Results}
\begin{figure}[htbp]
  \center
	\includegraphics[width=0.618\textwidth]{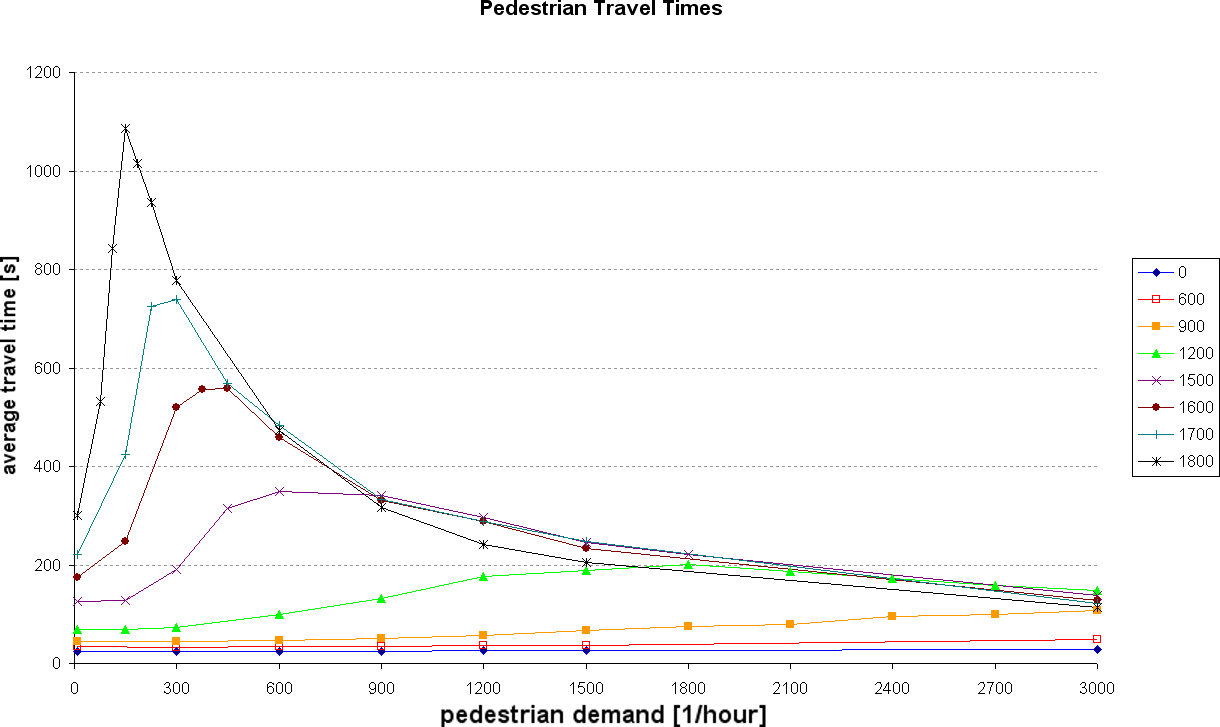}
	\caption{Dependency of pedestrians' travel times on pedestrian demand well below capacity and for different vehicle demands.}
	\label{fig:ptt1}
\end{figure}

\begin{figure}[htbp]
  \center
	\includegraphics[width=0.618\textwidth]{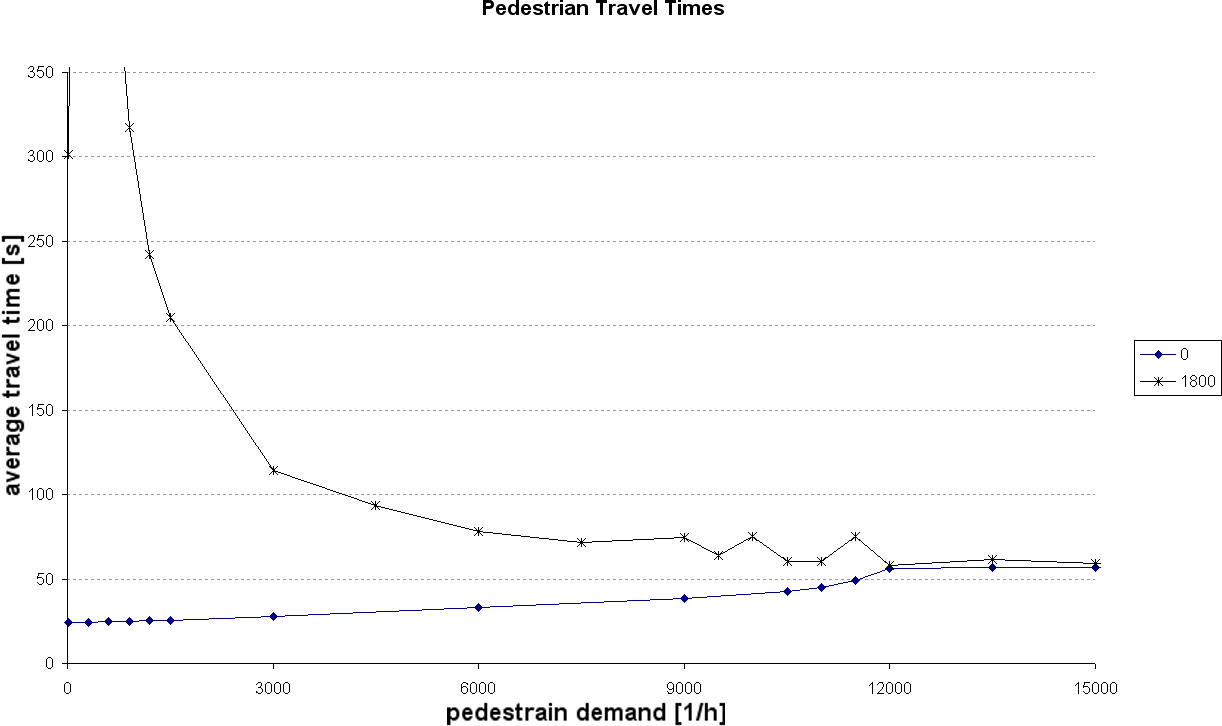}
	\caption{Dependency of pedestrians' travel times on pedestrian demand well below capacity and for the two extreme vehicle demands (none and 1,800/h). Note that in effect the simulations with a demand of 12,000 and 15,000 pedestrians (and above) are the same in all relevant aspects, as the density on the input area is quickly too large (above 5 pedestrians per sqm) to insert further pedestrians.}
	\label{fig:ptt2}
\end{figure}

\begin{figure}[htbp]
  \center
	\includegraphics[width=0.618\textwidth]{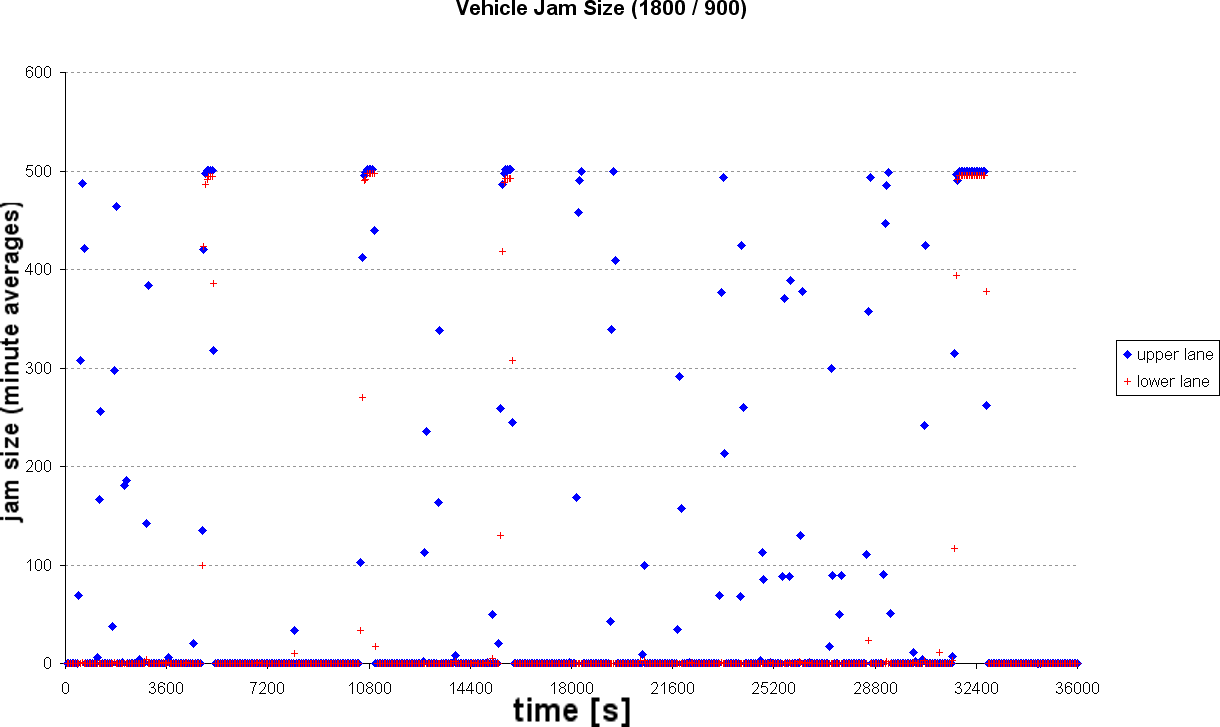} \\
	\vspace{12pt}
	\includegraphics[width=0.618\textwidth]{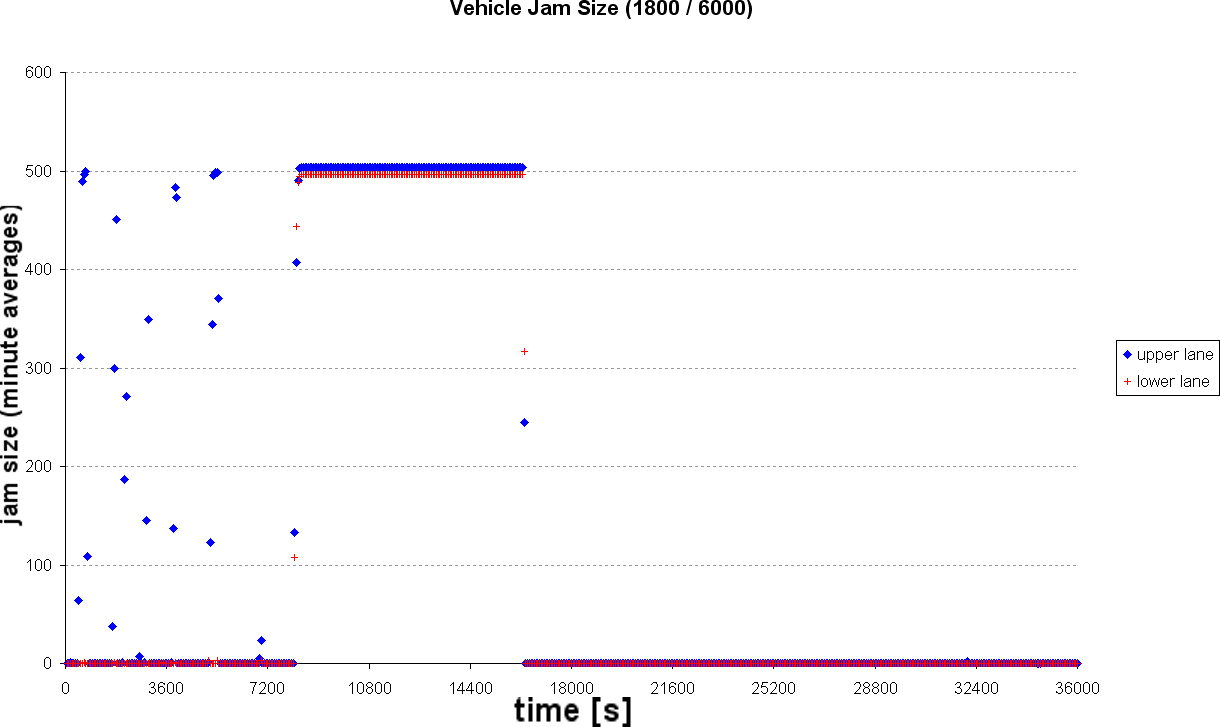} \\
	\caption{Vehicle jam size over time for a demand of 1,800 vehicles per hour and lane and 900 (upper) or 6,000 pedestrians per hour (lower).}
	\label{fig:vjss2}
\end{figure}

\begin{figure}[htbp]
  \center
	\includegraphics[width=0.618\textwidth]{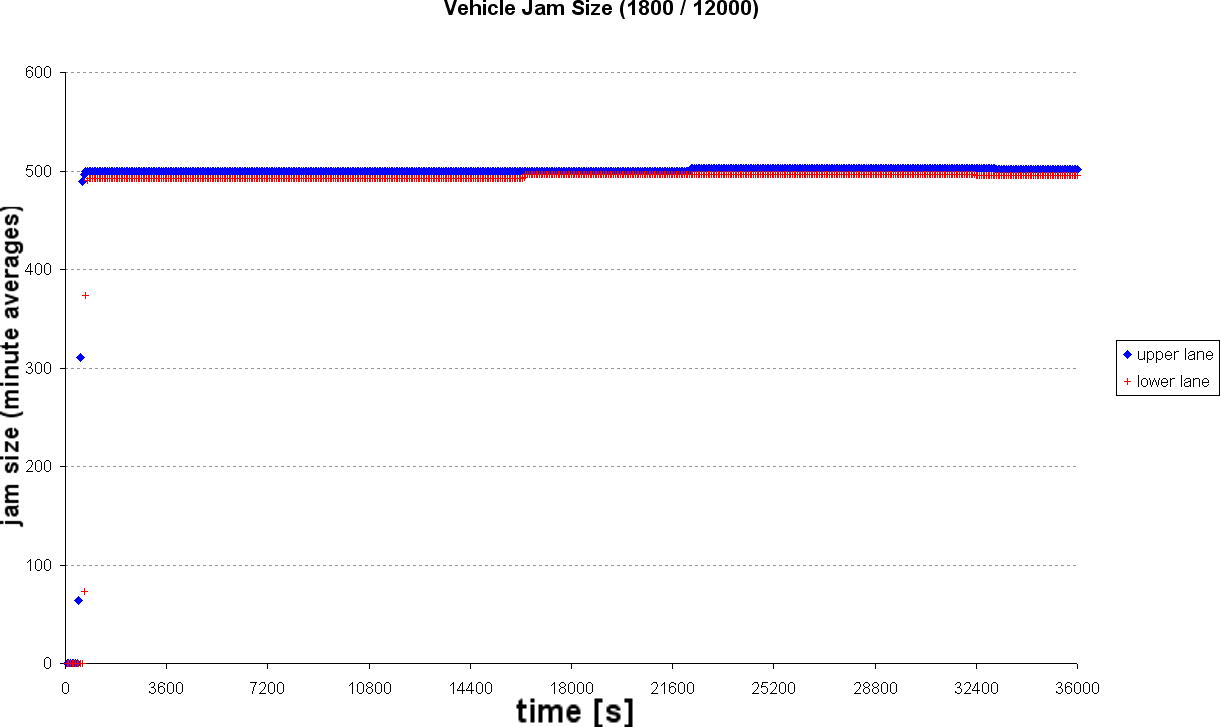} \\
	\vspace{12pt}
	\includegraphics[width=0.618\textwidth]{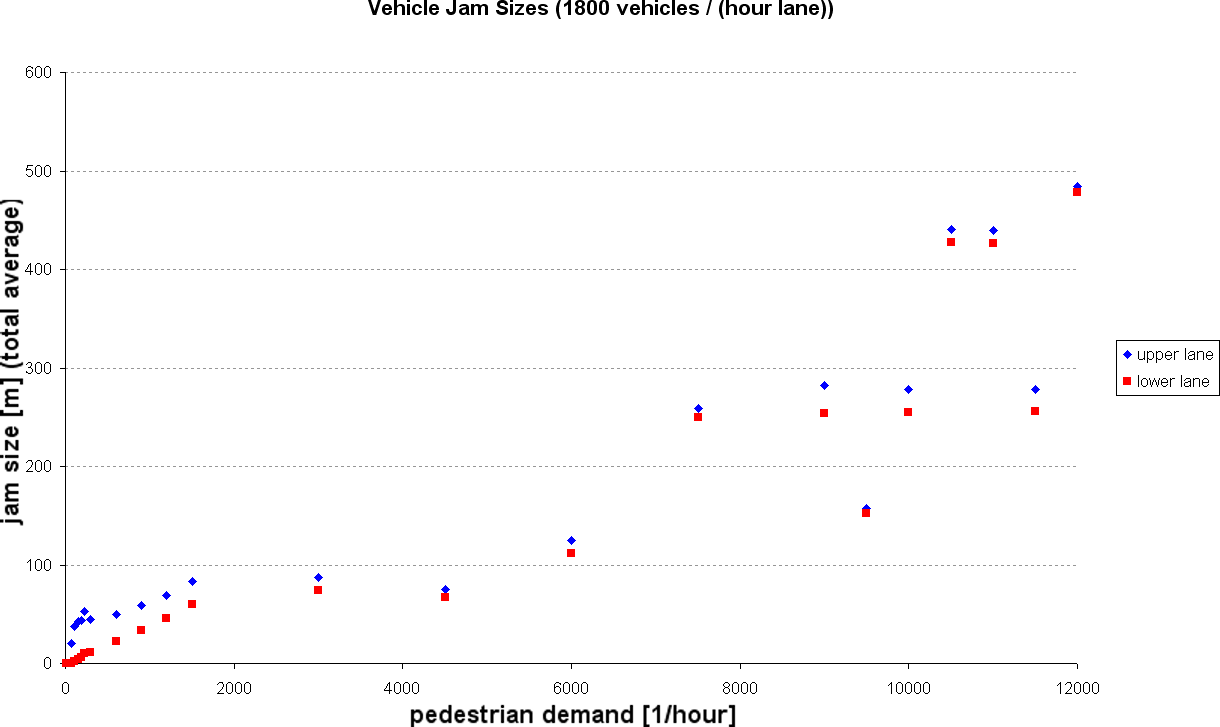} \\
	\caption{Upper: Vehicle jam size over time for a demand of 1,800 vehicles per hour and 12,000 pedestrians per hour. Lower: vehicle jam size time average in dependence of pedestrian demand.}
	\label{fig:vjss3}
\end{figure}

Figure \ref{fig:ptt1} shows the dependency of pedestrians' travel times on pedestrian demand for various vehicle demands. Above a vehicle demand threshold of about 700 to 800 vehicles per hour the dependencies show a local maximum. Above the maximum the flow of pedestrians is large enough that subsequent pedestrians can make a profit of vehicles having to slow down, when preceding pedestrians could not cross the street in time. Figure \ref{fig:ptt2} shows that the dependencies for no and large vehicle demand converge for pedestrian demand toward capacity. At pedestrian demands just below convergence the average jam length appears to be unstable, as the total simulation time is only a small multiple of the typical oscillation period (see figures \ref{fig:vjss2} and \ref{fig:vjss3}).

%
\nocite{_Bazzan2009,_Enzy2009}
\bibliographystyle{utphys_quotecomma}
\bibliography{027_TGF09}
%
\end{document}